\documentclass[lettersize,journal]{IEEEtran}
\usepackage{cite}
\usepackage{amsmath,amssymb,amsfonts,bm,amsthm}
\usepackage{graphicx}
\usepackage{caption}
\usepackage{subcaption}
\usepackage{textcomp}
\usepackage{dsfont}
\usepackage{xspace,mathtools}
\usepackage{bm}
\usepackage{multicol}
\usepackage{float}
\usepackage{stfloats}
\usepackage{optidef}
\usepackage[usenames, dvipsnames]{color}

\def\BibTeX{{\rm B\kern-.05em{\sc i\kern-.025em b}\kern-.08em
		T\kern-.1667em\lower.7ex\hbox{E}\kern-.125emX}}

\newcommand{\mat}[1]{\ensuremath{\mathbf{#1}}\xspace} 
\renewcommand{\vec}[1]{\ensuremath{\mathbf{#1}}\xspace} 

\DeclarePairedDelimiter\abs{\lvert}{\rvert}
\DeclarePairedDelimiter\norm{\lVert}{\rVert}

\newcommand{\sinr}{\textsf{SINR}}
\newcommand{\snr}{\textsf{SNR}}

\newcommand{\rcs}{\textsf{rcs}}
\newcommand{\ds}{\textsf{DS}}

\newcommand{\iui}{\textsf{IUI}}
\newcommand{\si}{\textsf{SI}}

\usepackage[compact]{titlesec}         
\usepackage[linesnumbered,ruled]{algorithm2e}
\newcounter{MYtempeqncnt}

\begin{document}
\title{Target Detection for OTFS-Aided Cell-Free MIMO ISAC System}

\author{
Shivani Singh, Amudheesan Nakkeeran, Prem Singh, Ekant Sharma, and Jyotsna Bapat
\thanks{Shivani Singh, Prem Singh, and Jyotsna Bapat are with the International Institute of Information Technology Bangalore, India (e-mail: shivani.singh@iiitb.ac.in; prem.singh@iiitb.ac.in; jbapat@iiitb.ac.in).
	
Amudheesan Nakkeeran is with the International Institute of Information Technology Bangalore, India, and IIITB COMET Foundation, India (e-mail: amudheesan.nakkeeran@iiitb.ac.in).
	
Ekant Sharma is with the Department of Electronics and Communication Engineering, Indian Institute of Technology Roorkee, India (e-mail: ekant@ece.iitr.ac.in).

This work has been submitted to the IEEE for possible publication. Copyright may be transferred without notice, after which this version may no longer be accessible.}
\vspace{-0.43cm}}



\maketitle

\begin{abstract}
This letter focuses on enhancing target detection performance for a multi-user integrated sensing and communication (ISAC) system using orthogonal time frequency space (OTFS)-aided cell-free multiple-input multiple-output (MIMO) technology in high-speed vehicular environments. We propose a sensing-centric (SC) approach for target detection using communication signals with/without sensing signals. Power allocation is optimized to maximize the sensing signal-to-noise ratio (SNR) of the proposed SC scheme while ensuring a required quality-of-service (QoS) for the communication user equipment (UEs), and adhering to each access point's (AP's) power budget. Numerical results show that the proposed SC scheme vastly outperforms a communication-centric method that minimizes the total power consumed at the APs subject to the same constraints.
\end{abstract}
\vspace{-0.2cm}

\section{Introduction}
Wireless sensing, including object detection and tracking, has traditionally been developed separately from cellular communication systems. However, the international mobile telecommunications (IMT)-2030 report recommends a shift to designing future networks with an integrated sensing and communication (ISAC) approach from the outset \cite{imt2030}. ISAC technology is expected to be crucial for 6G, integrating wireless communication and radar technologies to improve spectrum efficiency and reduce hardware costs. A key application of ISAC is in vehicle-to-everything (V2X)~technology \cite{chu2023integrated}. 

In parallel, cell-free multiple-input multiple-output (MIMO) has been gaining traction as a feasible candidate to realize ISAC systems \cite{mao2024communication, behdad2024multi}. The authors in~\cite{chu2023integrated} studied the integration of sensing and communication in cell-free massive MIMO systems employing orthogonal frequency division multiplexing (OFDM).
However, it is well known that OFDM waveform is not robust against the doubly dispersive channel experienced in high mobility applications like V2X. This necessitates alternative waveforms that are resilient towards high mobility scenarios. One promising alternative waveform is orthogonal time frequency space (OTFS) \cite{SaifK2021Derivation}, which multiplexes information symbols in the delay-Doppler (DD) domain to exploit time-invariant characteristics of the DD channel even in high-mobility environments. The OTFS can be implemented using the OFDM waveform with the additional use of pre- and post-processing 2D transforms~\cite{Yadav2024IRS-OTFS, ramachandran2020otfs}. 

The authors in \cite{mohammadi2022cell} considered cell-free massive MIMO with OTFS modulation and analyzed achievable downlink and uplink spectral efficiencies (SEs). Reference~\cite{das2024otfs} investigated an OTFS-aided full-duplex cell-free massive MIMO system to achieve high throughput gain in high mobility applications. Reference~\cite{VOFDM} focuses on cell-free ISAC systems and adopts vector OFDM (V-OFDM) as the integrated signal, to enable communication and detection of targets. The authors in~\cite{li2022novel} designed power allocation schemes for radar sensing in the ISAC transmission based on the spatially spread OTFS. 

The existing literature is yet to investigate cell-free and OTFS technologies for ISAC system design, an exception being the work in~\cite{fan2024powerallocationcellfreemassive}, 
 which derived the downlink SE by considering multi-antenna access points (APs) and optional sensing beams. They employed maximal ratio precoding for communication signals and conjugate beamforming for the sensing beams, aiming to maximize the minimum communication signal-to-interference-plus-noise ratio (SINR) subject to a constraint on sensing performance. To the best of our knowledge, ours is the first study to investigate target detection performance in OTFS-aided cell-free MIMO ISAC systems 
with the following contributions.
\begin{itemize}
    \item For an OTFS-aided cell-free massive MIMO ISAC system, we design regularized zero-forcing precoders for the communication users and employ nullspace beamforming for sensing target detection to minimize interference with the communication signals.
    \item We propose a sensing-centric (SC) scheme for target detection and quantify its performance by deriving a maximum a-posteriori ratio test (MAPRT) detector.  We maximize the target detection probability of the proposed method by designing a power allocation algorithm that maximizes the sensing signal-to-noise ratio (SNR) while meeting quality-of-service (QoS) constraints for the communication users and the power budget of the transmit access points (APs). 
    \item Our simulation results evaluate the target detection performance of the proposed SC scheme using communication signals, and then both communication and sensing signals. To benchmark the proposed SC scheme, we compare its performance with a communication-centric approach that minimizes total AP power consumption. Our results show the target detection probability of the communication-centric scheme is close to zero even when the radar cross section (RCS) variance of the target is  $-20$ dBsm (decibels relative to one square meter), while the proposed SC scheme leads to target detection probability of $1$ for the same value of RCS.  
\end{itemize}

\section{SYSTEM MODEL}
We consider a cell-free MIMO ISAC-OTFS system as shown in Fig. 1, facilitating downlink communication and multi-static sensing (transmitter and receiver not co-located). The system model consists of $N_{tx}$ and $N_{rx}$ number of transmit and receive APs respectively, each equipped with $L$ antennas in a uniform linear array (ULA). The APs are distributed in a geographical area and are connected via fronthaul links to a cloud radio access network (C-RAN) architecture, which enables joint signal processing ~\cite{mohammadi2022cell}. The transmit APs jointly serve $N_{ue}$ number of single-antenna UEs and detect a target by transmitting integrated signals consisting of communication signals and an optional sensing signal on the same time-frequency resources. The receive APs collect the echo signals to jointly detect a target at a specified location. We assume a vehicular environment between the APs and UEs, which results in a doubly dispersive (time and frequency selective) wireless channel~\cite{das2024otfs}. For reliable transmission over this channel, the OTFS waveform is used by the ISAC system. We next describe the transmitter of the proposed cell-free massive MIMO ISAC-OTFS system.
\begin{figure}[t] 
	\begin{center}
		\includegraphics[scale=0.3]{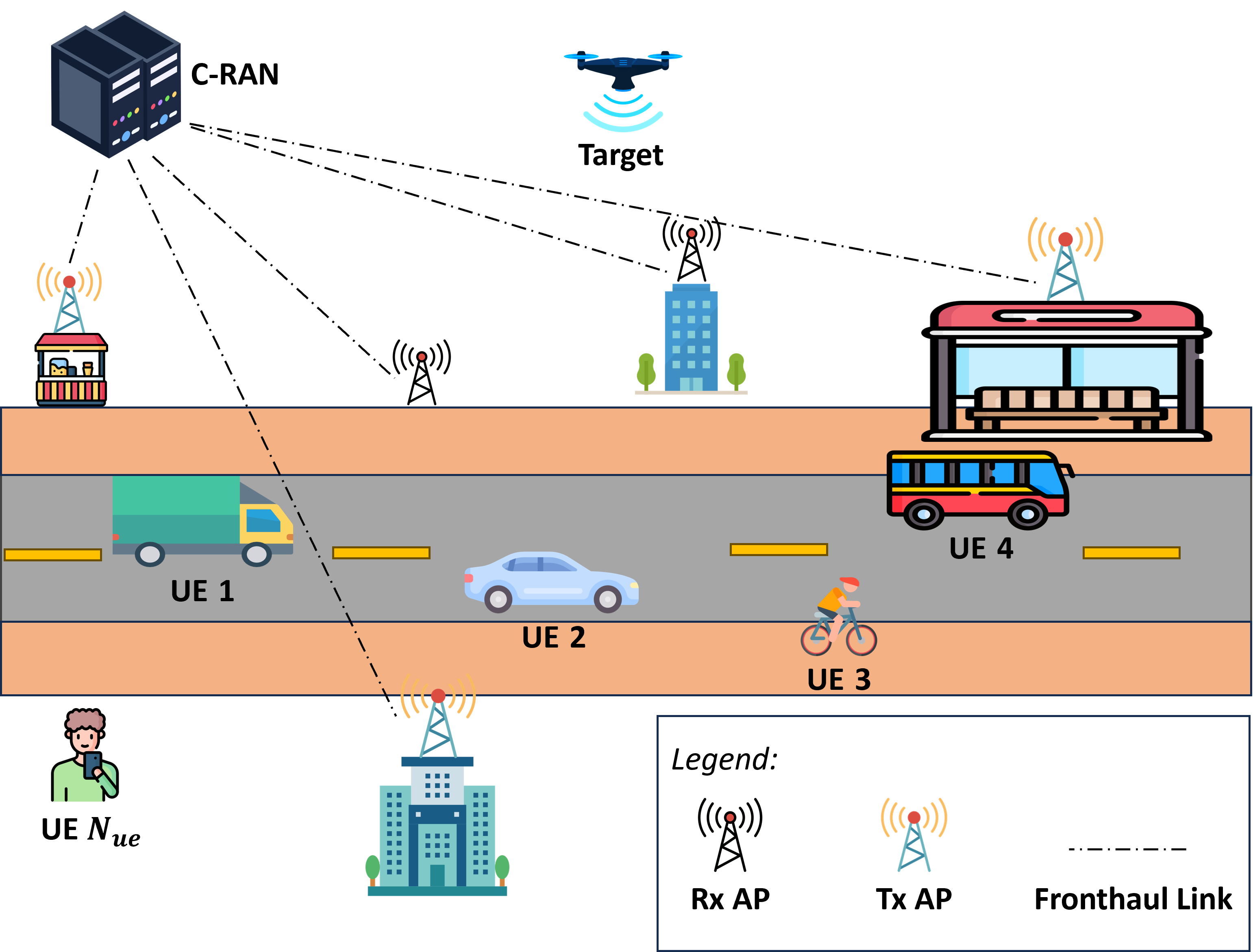}
	\end{center}
	\caption{\small Illustration of a cell-free MIMO ISAC-OTFS system.}
 \label{Fig:JCAS}
 \vspace{-0.5cm}
\end{figure}

\subsection{ISAC-OTFS Transmitter Structure:} The frame of the OTFS waveform is assumed to have $M$ subcarriers (delay bins) and $N$ symbols (Doppler bins). Let $\mat{V}_u \in \mathbb{C}^{M \times N}$ be a delay-Doppler (DD) domain transmit symbol matrix for UE $u$. After using the inverse symplectic finite Fourier transform (ISFFT) and Heisenberg transform  at the transmit APs, the time domain transmit symbol vector $\vec{x}_u\in \mathbb{C}^{MN \times 1}$ for UE $u, 1 \leq u \leq N_{ue}$, is obtained as\cite{Singh2022Low-Complexity}
\begin{equation} \label{x0}
	\vec{x}_u = \text{vec}(\mat{F}_{M}^H\mat{F}_M\mat{V}_u\mat{F}_{N}^H) = (\mat{F}_{N}^H \otimes \mat{I}_M)\vec{v}_u,
\end{equation}
where $\vec{v}_u = \text{vec}(\mat{V}_{u})$ and $\mat{F}_M$ is the $M \times M$ discrete Fourier transform matrix. Let $\vec{x}_0$ denote an optional sensing signal which is independent of communication signals and is generated in the same way as \eqref{x0}. The transmit signal $\vec{d}_k \in \mathbb{C}^{LMN \times 1}$ for all the UEs, along with the sensing signal, from AP $k$ is given as 
\begin{equation} \label{dk}
	\vec{d}_k = \sum_{u=0}^{N_{ue}} \sqrt{\eta_u}\mat{W}_{u,k}^H\vec{x}_u = \mat{W}_k^H\mat{D}\boldsymbol{\eta},
\end{equation}
where $\mat{W}_k = \big[ \mat{W}_{0,k}^T,~ \mat{W}_{1,k}^T,~ \dots,~ \mat{W}_{N_{ue},k}^T \big]^T $ is the concatenated precoding matrix of size $ (N_{ue}+1)MN \times LMN$, $\boldsymbol{\eta} = [\sqrt{\eta_0},~ \sqrt{\eta_1},~ \dots,~ \sqrt{\eta_{N_{ue}}}]^T $ is the power control coefficient vector, and $\mat{D}=\text{blkdiag}[\vec{x}_{0},\vec{x}_{1},\ldots, \vec{x}_{N_{ue}}]$ is a block diagonal matrix of size $MN(N_{ue}+1) \times (N_{ue}+1)$ obtained from the sensing and communication symbols.
The average transmit power of AP $k$ is computed from \eqref{dk} as
\begin{equation} \label{APpower}
    P_k ~=~ \mathbb{E}\left[ \lVert \vec{d}_k \rVert^2 \right] = \sum_{u=0}^{N_{ue}} \eta_u \norm*{\mat{W}_{u,k}}_{F}^2.
\end{equation}

\subsection{ISAC-OTFS Channel Model and Precoder Design}
The DD domain communication channel $\mat{H}_{u,k}^{(l)} \in \mathbb{C}^{MN \times MN}$ between antenna $l$ of AP $k$ and UE $u$ in matrix form is given as \cite{li2022novel}
\begin{equation} \label{DD-channel}
	\mat{H}_{u,k}^{(l)} = \sum_{p=1}^{P} h_{u,k}^{(l,p)} e^{ \left(j \pi (l-1) \sin \left( \theta_{u,k}^{(p)} \right) \right) } \boldsymbol{\Pi}^{m_k^{(p)}} \boldsymbol{\Delta}^{n_k^{(p)}},
\end{equation}
where $P$ is the number of DD domain propagation paths, $h_{u,k}^{(l,p)}$ models the channel gain for path $p$ between antenna $l$ of AP $k$ and UE $u$, $\theta_{u,k}^{(p)}$ is the angle of departure from AP $k$ towards UE $u$ through path $p$, and $m_k^{(p)}$ and $n_k^{(p)}$ respectively denote the delay and Doppler taps for the path $p$. In (\ref{DD-channel}), the matrices $\boldsymbol{\Pi} = \text{circ}\big\lbrace \left[ 0,1,\dots,0 \right]_{MN \times 1}^T \big\rbrace$, and $\boldsymbol{\Delta} = \text{diag}\left\lbrace \left[ 1, e^{j2\pi / MN}, \dots, e^{j2\pi(MN-1) / MN} \right] \right\rbrace$ characterize the delay and Doppler shifts respectively.

It is imperative to design precoders for ISAC systems carefully in order to not only reduce the interference between the communication signals intended for different UEs but also to nullify the interference caused by the sensing signal to the communication signals. Thanks to C-RAN and the centralized operation in a cell-free MIMO system, joint selection of precoding vectors for each UE and the target is possible. To obtain the same, we define $\mat{H}_{u,k} = \big[ \mat{H}_{u,k}^{(1)},~ \mat{H}_{u,k}^{(2)},~ \dots,~ \mat{H}_{u,k}^{(L)} \big]$ and concatenate the $\mat{H}_{u,k}$ matrices to get $\mat{H}_u = \left[ \mat{H}_{u,1},~ \mat{H}_{u,2},~ \dots,~ \mat{H}_{u,N_{tx}} \right]$ as the DD domain communication channel from all the $N_{tx}L$ transmit antennas to UE $u$.
We then construct the unit-norm regularized zero forcing (RZF) precoder for UE $u$ as  
\begin{equation}
    \mat{W}_{u} = \frac{\left[ \sum_{v=1}^{N_{ue}} \mat{H}_{v} \mat{H}_{v}^H + \psi\mat{I}_{MN} \right]^{-1} \mat{H}_{u}}{\left\Vert \left[ \sum_{v=1}^{N_{ue}} \mat{H}_{v} \mat{H}_{v}^H + \psi\mat{I}_{MN} \right]^{-1} \mat{H}_{u} \right\Vert_{F}},
\end{equation}
where $\psi$ is the regularization parameter and $\left\Vert \cdot \right\Vert_{F}$ is the Frobenius norm. The RZF precoder is chosen for balancing between interference suppression and having strong desired signal powers. The precoders for UE $u$ at the individual transmit APs is extracted from $\mat{W}_{u}$ as $\mat{W}_{u} = \left[ \mat{W}_{u,1}, \mat{W}_{u,2},~ \dots,~ \mat{W}_{u,N_{tx}} \right]$. 
When dedicated sensing symbols are used by the transmit APs, we construct a sensing precoder, which is in the nullspace of the subspace formed by all the UE channel matrices, as
\begin{equation}
    \mat{W}_{0} = \frac{\vec{a} \otimes \left( \mat{I}_{MN} - \mat{H} \left( \mat{H}^H\mat{H} \right)^\dag \mat{H}^H \right)}{\left\Vert \vec{a} \otimes \left( \mat{I}_{MN} - \mat{H} \left( \mat{H}^H\mat{H} \right)^\dag \mat{H}^H \right) \right\Vert_{F}},
\end{equation}
where $\mat{H} = \left[ \mat{H}_{1},~ \mat{H}_{2},~ \dots,~ \mat{H}_{N_{ue}} \right]^T$ is the concatenated channel seen by all the UEs, and $\vec{a} = \left[ \vec{a}\left( \varphi_1, \vartheta_1 \right)^T,~ \vec{a}\left( \varphi_2, \vartheta_2 \right)^T,~ \dots,~ \vec{a}\left( \varphi_{N_{tx}}, \vartheta_{N_{tx}} \right)^T \right]^T$ is the antenna array response vector from all transmit APs towards the target. Here, $\varphi_k$ and $\vartheta_k$ are the azimuth and elevation angles respectively for transmit AP $k$. The precoder $\mat{W}_{0}$ ensures that the sensing signal does not interfere with the communication signals intended for the UEs. The sensing precoders at the individual transmit APs can be then extracted from $\mat{W}_{0}$.

\subsection{ISAC-OTFS Receiver Structure} 
The $MN \times 1$ received signal at UE $u$ from all the transmit APs is $\vec{y}_u= \sum_{k=1}^{N_{tx}} \mat{H}_{u,k} \vec{d}_k + \vec{n}_u$, which using \eqref{dk},~becomes
\begin{align}\label{yu}
    \vec{y}_u &= \underbrace{\sqrt{\eta_u} \mat{H}_{u} \mat{W}_{u}^H\vec{x}_u}_{\text{Desired signal}} ~+~ \underbrace{\sqrt{\eta_0} \mat{H}_{u} \mat{W}_{0}^H\vec{x}_0}_{\text{Interference from sensing symbol}} \nonumber \\
    	&  +~ \underbrace{\sum_{v=1, v\neq u}^{N_{ue}} \sqrt{\eta_v} \mat{H}_{u} \mat{W}_{v}^H\vec{x}_v}_{\text{Interference from the other UEs}} ~+~ \underbrace{\vec{n}_u}_{\text{Noise}}, 
\end{align}
where $\vec{n}_{u} \sim \mathcal{N}_{\mathbb{C}} \left( 0, \sigma_n^2 \mat{I}_{MN} \right)$ is the noise vector at the UE $u$. We observe that the UE $u$ gets interference not only from other users but also from the sensing symbols. 

The $N_{rx}$ receive APs in Fig. \ref{Fig:JCAS} collect signals sent by all the $N_{tx}$ transmit APs and their reflections from the target. Since the APs can jointly process their signals via the C-RAN architecture, the receive APs can remove the target-free components from their received signal. Then, the received signal $\vec{y}_r \in \mathbb{C}^{LMN \times 1}$ at the receive AP $r$ through the target (when present) is 
\begin{equation}\label{yr_1}
	\vec{y}_r = \sum_{k=1}^{N_{tx}} \xi_{r,k} \underbrace{\left( \vec{a}(\phi_r, \theta_r) \vec{a}^T(\varphi_k, \vartheta_k) \otimes \mat{H}_{r,k} \right) \vec{d}_k}_{\triangleq ~\boldsymbol{\omega}_{r,k}} + \underbrace{\vec{n}_r}_{\text{Noise}},
\end{equation}
where $\mat{H}_{r,k}$ is the time delay domain \cite{li2022novel} equivalent of the DD domain channel matrix between receive AP $r$ and transmit AP $k$, $\xi_{r,k} \sim \mathcal{N}_{\mathbb{C}} \left(0, \sigma_{\rcs}^2\right)$ is the channel gain of the reflected path through the target between transmit AP $k$ and receive AP $r$ including the RCS of the target, and $\vec{n}_{r} \sim \mathcal{N}_{\mathbb{C}} \left( 0, \sigma_n^2 \mat{I}_{LMN} \right)$ is the noise vector at receive AP $r$.  Concatenating $\boldsymbol{\omega}_{r,k}$ and $\xi_{r,k}$ in \eqref{yr_1} as $\mat{\Omega}_r = \left[ \boldsymbol{\omega}_{r,1}, \boldsymbol{\omega}_{r,2}, \dots, \boldsymbol{\omega}_{r,N_{tx}} \right]$ and  $\boldsymbol{\xi}_r = \left[ \xi_{r,1}, \xi_{r,2}, \dots, \xi_{r,N_{tx}} \right]^T$, we rewrite \eqref{yr_1} as
\begin{equation} \label{yr}
    \vec{y}_r = \mat{\Omega}_r \boldsymbol{\xi}_r + \vec{n}_r.
\end{equation}
Now, stacking the vectors $\vec{y}_r$ from all the receive APs as $\vec{y} = \left[ \vec{y}_1^T, \vec{y}_2^T, \dots, \vec{y}_{N_{rx}}^T \right]^T$, we can compactly represent $\vec{y}$ as
\begin{equation}\label{rxsig}
	\vec{y} = \mat{\Omega} \boldsymbol{\xi} + \vec{n},
\end{equation}
where the matrix $\mat{\Omega} = \mbox{blkdiag} \left\lbrace \mat{\Omega}_1, \mat{\Omega}_2, \dots, \mat{\Omega}_{N_{rx}} \right\rbrace$, the vectors $\boldsymbol{\xi} = \left[ \boldsymbol{\xi}_1^T, \boldsymbol{\xi}_2^T, \dots, \boldsymbol{\xi}_{N_{rx}}^T \right]^T$, and $\vec{n} = \left[ \vec{n}_1^T, \vec{n}_2^T, \dots, \vec{n}_{N_{rx}}^T \right]^T$. If the target is not present, there are no reflections from it, and thus $\mat{\Omega}$ is equal to \mat{0}. 


\subsection{ISAC-OTFS MAPRT Detector} \label{maprt}
\begin{figure*}[!t]
\setcounter{MYtempeqncnt}{\value{equation}}
\setcounter{equation}{13}
\begin{equation} \label{dec_rule}
	\left\lbrace\ln\left[\bigg( \frac{1}{ \pi\sigma_{\rcs}^2} \bigg)^{N_{tx}N_{rx}}\right] + \sigma_{n}^{-2} \left( \vec{y}^H\mat{\Omega} \right) \times \left( \mat{\Omega}^H \mat{\Omega} + \sigma_n^2 \left( \sigma_{\rcs}^2 \times \mat{I}_{N_{tx}N_{rx}} \right)^{-1} \right)^{-1} \left( \mat{\Omega}^H\vec{y} \right)\right\rbrace \begin{matrix}
    \mathcal{H}_1\\[-1mm]>\\[-2.5mm] <\\[-1mm] \mathcal{H}_0
    \end{matrix} \lambda.
\end{equation}
\setcounter{equation}{\value{MYtempeqncnt}}
\hrulefill
\end{figure*}
We use the probability of target detection $P_d$, defined as the conditional probability of detecting the target given that the target is present, as the measure for sensing performance. We use a MAPRT target detector \cite{behdad2024multi} based on the power allocation strategy to be explained in Section \ref{powerAlloc}. The binary hypothesis problem in the MAPRT detector is formed using \eqref{rxsig} as 
\begin{equation}\label{MPART_Model}
	\mathcal{H}_0 : \vec{y} = \vec{n}, \quad \mathcal{H}_1 : \vec{y} = \mat{\Omega} \boldsymbol{\xi} + \vec{n},
\end{equation}
where $\mathcal{H}_0$ is the null hypothesis representing the case when the target is absent and $\mathcal{H}_1$ represents the presence of the target. The noise vector $\vec{n}\sim \mathcal{N}_{\mathbb{C}} \left( 0, \sigma_n^2 \mat{I}_{N_{rx}LMN} \right)$. We form the joint RCS estimation and hypothesis testing as $$\left( \hat{\boldsymbol{\xi}}, \hat{\mathcal{H}} \right) = \arg\max_{\boldsymbol{\xi}, \mathcal{H}} f(\boldsymbol{\xi}, \mathcal{H} \vert \vec{y}),$$where $\mathcal{H} = \left\lbrace \mathcal{H}_0, \mathcal{H}_1 \right\rbrace$, and $f(\boldsymbol{\xi}, \mathcal{H} \vert \vec{y})$ is the conditional joint probability density functions (pdfs) of $\boldsymbol{\xi}$ and $\mathcal{H}$ given $\vec{y}$. The corresponding MAPRT decision rule is then
\begin{equation} \label{maprt_orig}
    \max_{\boldsymbol{\xi}} ~\frac{ f\left( \vec{y} | \boldsymbol{\xi}, \mathcal{H}_1 \right)f\left( \boldsymbol{\xi} | \mathcal{H}_1 \right)}{f \left( \vec{y} | \mathcal{H}_0 \right)} \begin{matrix}
    \mathcal{H}_1\\[-1mm]>\\[-2.5mm] <\\[-1mm] \mathcal{H}_0
    \end{matrix} \lambda,
\end{equation}
where $\lambda$ is the target detection threshold. The conditional pdfs in \eqref{maprt_orig} are calculated using \eqref{rxsig} and \eqref{MPART_Model} as
\begin{equation} \label{maprt_den}
    f \left( \vec{y} | \mathcal{H}_0 \right) =  \left( \frac{1}{\pi \sigma_n^2} \right)^{LMNN_{rx}} \exp\left( -\frac{\| \vec{y} \|^2}{\sigma_n^2} \right).
\end{equation} \setcounter{MYtempeqncnt}{\value{equation}} We similarly calculate the pdfs $f\left( \vec{y} | \boldsymbol{\xi}, \mathcal{H}_1 \right)$ and $f\left( \boldsymbol{\xi} | \mathcal{H}_1 \right)$ using \eqref{rxsig} and \eqref{MPART_Model} and substitute them in \eqref{maprt_orig} to derive the MAPRT decision rule as shown in \eqref{dec_rule} at the top of this page.
\setcounter{equation}{\value{MYtempeqncnt}+1}
We then optimize the detection performance of the MAPRT detector for the proposed SC scheme by designing an optimal power allocation algorithm described next.

\section{Power Allocation for ISAC-OTFS System} \label{powerAlloc}
We begin by calculating the communication SINR and sensing SNR before proceeding to the power allocation strategy. From (\ref{yu}), we calculate the SINR for UE $u$ as 
\begin{equation} \label{ue_sinr}
	\sinr_u = \frac{\eta_u\abs*{\ds_u}^2}{\sum_{\substack{v=1, \\ v \neq u}}^{N_{ue}} \eta_v\abs*{\iui_{u,v}}^2 ~+~ \eta_0\abs*{\si_u}^2 ~+~ MN\sigma_{n}^2},
\end{equation}
where $\ds_u = \left\Vert \mat{H}_{u} \mat{W}_{u}^H\vec{x}_u \right\Vert_{F},~ \iui_{u,v} = \left\Vert\mat{H}_{u} \mat{W}_{v}^H\vec{x}_v \right\Vert_{F}$ and $\si_u = \left\Vert \mat{H}_{u} \mat{W}_{0}^H\vec{x}_0 \right\Vert_F$ are the desired signal power at UE $u$, inter-user interference power at UE $u$ caused by UE $v$, and the interference power at UE $u$ due to sensing respectively.

The sensing SNR for the received signal $\vec{y}$ in (\ref{rxsig}) is 
\begin{equation}\label{sensingSNR}
	\snr_\textsf{sensing} = \frac{\mathbb{E} \left[ \norm{\mat{\Omega} \boldsymbol{\xi}}^2 \right]}{\mathbb{E} \left[ \norm{\vec{n}}^2 \right]} = \frac{\mathbb{E} \left[ \norm{\mat{\Omega} \boldsymbol{\xi}}^2 \right]}{LMNN_{rx}\sigma_{n}^2},
\end{equation}
where $\mathbb{E} \left[ \norm{\mat{\Omega} \boldsymbol{\xi}}^2 \right]$ is calculated using \eqref{dk} and \eqref{yr} as
\begin{equation} \label{sensingPower}
	\mathbb{E} \left[ \norm{\mat{\Omega} \boldsymbol{\xi}}^2 \right] = \sum_{r=1}^{N_{rx}} \sum_{k=1}^{N_{tx}} \mathbb{E} \left[ \boldsymbol{\omega}_{r,k}^H\xi_{r,k}^* \xi_{r,k}\boldsymbol{\omega}_{r,k} \right] = \boldsymbol{\eta}^T \mat{\Psi} \boldsymbol{\eta}.
\end{equation}
Here the matrix
\begin{align} \label{Amat}
	\mat{\Psi} &= \mat{D}^H \left( \sum_{r=1}^{N_{rx}} \sum_{k=1}^{N_{tx}} \sigma_{\rcs}^2 \mat{W}_k \left( \vec{a}^*(\varphi_k, \vartheta_k) \vec{a}^H(\phi_r, \theta_r) \right. \right. \nonumber \\
	& \quad \left. \times \quad   \vec{a}(\phi_r, \theta_r) \vec{a}^T(\varphi_k, \vartheta_k) \otimes \mat{I}_{MN} \right) \mat{W}_k^H \Biggr) \mat{D}.
\end{align}
Using \eqref{sensingPower} in \eqref{sensingSNR}, we write the sensing \snr~as
\begin{equation} \label{snr_sensing}
	\snr_\textsf{sensing} = \frac{1}{LMNN_{rx}\sigma_n^2} \times \boldsymbol{\eta}^T \mat{\Psi} \boldsymbol{\eta}.
\end{equation}
We now design a power allocation algorithm for the MAPRT detector. For this, we maximize the sensing SNR given that the target is present while ensuring that the QoS constraints for the communication UEs and the power budget constraints of each transmit AP are met. Thus, the sensing-centric optimization problem is cast as
\begin{maxi!}{\boldsymbol{\eta} \geq \vec{0}}{\snr_\textsf{sensing}}{\label{opti}}{}
\addConstraint{\sinr_u} {\geq \gamma_{\textsf{thresh}}}, \quad 1 \leq u \leq N_{ue} \label{uethresh}
\addConstraint{P_k}{\leq P_\textsf{max}}, \quad 1 \leq k \leq N_{tx} \label{budget},
\end{maxi!}
where $\gamma_{\textsf{thresh}}$ is the minimum SINR required to satisfy the communication QoS, and $P_\textsf{max}$ is the power budget per AP.
The objective function in \eqref{opti} is expressed using \eqref{snr_sensing}, where $\mat{\Psi}$ is a symmetric Hermitian matrix. As a result, the term $\boldsymbol{\eta}^T \mathcal{I}\left\lbrace \mat{\Psi} \right\rbrace \boldsymbol{\eta}$ involving the imaginary part of $\mat{\Psi}$ is zero. Therefore, we only need to consider the real part, $\boldsymbol{\eta}^T \mathcal{R}\left\lbrace \mat{\Psi} \right\rbrace \boldsymbol{\eta}$. Additionally, the constant multiplicative factor $\frac{1}{LMNN_{rx}\sigma_n^2}$ in \eqref{snr_sensing} does not influence the optimal solution and can be omitted without loss of generality.
Thus, we focus simply on maximizing $\boldsymbol{\eta}^T \mathcal{R}\left\lbrace \mat{\Psi} \right\rbrace \boldsymbol{\eta}$, or equivalently, minimizing $-\boldsymbol{\eta}^T \mathcal{R}\left\lbrace \mat{\Psi} \right\rbrace \boldsymbol{\eta}$. 

The constraints in \eqref{uethresh} are non-convex, so we rewrite them as second-order cone (SOC) constraints using \eqref{ue_sinr} as 
\begin{align*}
    &\frac{\eta_u \abs*{\ds_u}^2}{\gamma_{\textsf{thresh}}} \geq \sum_{\substack{v=1, \\ v \neq u}}^{N_{ue}} \eta_v\abs*{\iui_{u,v}}^2 ~+~ \eta_0\abs*{\si_u}^2 ~+~ MN\sigma_{n}^2 \nonumber \\
    & \iff \abs*{\ds_u}\sqrt{\frac{\eta_u}{\gamma_{\textsf{thresh}}}} \geq \nonumber \\
    & \quad\quad\quad\quad \sqrt{\eta_0\abs*{\si_u}^2 + \sum_{v=1, v \neq u}^{N_{ue}} \eta_v\abs*{\iui_{u,v}}^2 + MN\sigma_{n}^2},
\end{align*}
which is succinctly written for $1 \leq u \leq N_{ue}$ as
\begin{equation} \label{soc-uethresh}
    \norm*{ \begin{bmatrix}
        \mat{A}_u & \vec{0} \\
        \vec{0}^T & \sigma_n \sqrt{MN}
    \end{bmatrix} \begin{bmatrix}
        \boldsymbol{\eta} \\ 1
    \end{bmatrix} } ~\leq~ \abs*{\ds_u}\sqrt{\frac{\eta_u}{\gamma_{\textsf{thresh}}}}.
\end{equation}
Here $\mat{A}_u \!= \!\text{diag} \left\lbrace \left[ \abs*{\si_u},~ \abs*{\iui_{u,1}},~ \dots,~ \abs*{\iui_{u,u-1}},~ 0 \right.\right.$, $\left.\left.\abs*{\iui_{u,u+1}},~ \dots,~ \abs*{\iui_{u,N_{ue}}} \right] \right\rbrace$. Similarly, we rewrite the constraints in \eqref{budget} as SOC constraints using \eqref{APpower} as
\begin{align} \label{soc-budget}
   \sqrt{\sum_{u=0}^{N_{ue}} \eta_u \norm*{\mat{W}_{u,k}}_{F}^2} &\leq \sqrt{P_{\textsf{max}}} \nonumber \\
   \iff \norm*{\mat{G}_k^{T} \boldsymbol{\eta}} ~&\leq~ \sqrt{P_\textsf{max}}, \quad 1 \leq k \leq N_{tx},
\end{align}
where $\mat{G}_k = \text{diag} \left\lbrace \left[ \norm*{\mat{W}_{0,k}}_F, \norm*{\mat{W}_{1,k}}_F, \dots, \norm*{\mat{W}_{N_{ue},k}}_F \right] \right\rbrace$. Thus, the optimization problem in (\ref{opti}) is rewritten as
\begin{mini!}{\boldsymbol{\eta} \geq \vec{0}}{-\boldsymbol{\eta}^T \mathcal{R}\{\mat{\Psi}\} \boldsymbol{\eta} }{\label{ccopti}}{}
	\addConstraint{(\ref{soc-uethresh}), (\ref{soc-budget})}
\end{mini!}
We solve the above optimization problem using the concave-convex procedure in Algorithm \ref{alg:cap}, which starts with a random feasible power allocation vector $\boldsymbol{\eta}^{(0)}$ and solves a sequence of convex optimization problems to generate progressively better iterates $\boldsymbol{\eta}^{(t)}$ until a stopping condition is met. This concave-convex approach is shown to converge in \cite{CCP}.
\begin{algorithm}[t]
\DontPrintSemicolon 
\textbf{Input:} A feasible initial iterate $\boldsymbol{\eta}^{(0)}$, tolerance $\epsilon > 0 $\;
\textbf{Initialize:} Set iteration counter to $t=0$\;
  \Repeat{$\|\mat{\Psi}\left(\boldsymbol{\eta}^{(t)}-\boldsymbol{\eta}^{(t-1)}\right)\| \leq \epsilon$} {
    Set $t := t+1$ \;
    Generate $\boldsymbol{\eta}^{(t)}$ as the solution to:
\begin{mini!}{\boldsymbol{\eta} \geq \vec{0}}{-\boldsymbol{\eta}^T \mathcal{R}\{\mat{\Psi}\} \boldsymbol{\eta}^{(t-1)} }{\label{ccopti}}{}
	\addConstraint{(\ref{soc-uethresh}), (\ref{soc-budget})}
\end{mini!}\;
  }
\Return: {The transmit power coefficients $\boldsymbol{\eta}^{(t)}.$}\;
\caption{\small ISAC-OTFS Power Allocation Algorithm}
\label{alg:cap}
\end{algorithm}


\section{Numerical Results}
We simulate the results for a cell-free MIMO ISAC-OTFS setup in which AP locations are generated randomly using uniform distribution once in a 500 m $\times$ 500 m area. The $N_{rx}$ closest APs to the target, which appears at the center of the area, are designated as the receive APs. We consider $N_{tx} = 8$ transmit APs, $N_{rx} = 2$ receive APs, and $N_{ue} = 2$ UEs unless stated otherwise. Each AP has $L = 2$ antennas and maximum transmit power $P_\textsf{max} = 1$ Watt. The minimum threshold to satisfy communication QoS is $\gamma_{\textsf{thresh}} = 2$ dB. The MAPRT detector threshold $\lambda$ is chosen by setting the false alarm probability $\left( P_{fa} \right)$ to be $10^{-5}$. The carrier frequency and noise power are chosen to be 1.9 GHz and $-94$ dBm respectively. For the OTFS waveform, we choose $(M,N) = (16,8)$ and $P = 5$ DD domain channel paths between the transmit APs and UEs. The minimum and maximum delay spreads are 2.08 $\mu s$ and 10.41 $\mu s$ respectively. The minimum and maximum Doppler spreads are $0$ Hz and $1.88$ KHz respectively which correspond to relative velocity ranging from 0 kmph to 500 kmph approximately \cite{ramachandran2020otfs}. The results are averaged across $100$ UE positions. 

\begin{figure*}
     \centering
     \begin{subfigure}[b]{0.329\textwidth}
         \centering
         \includegraphics[width=\textwidth]{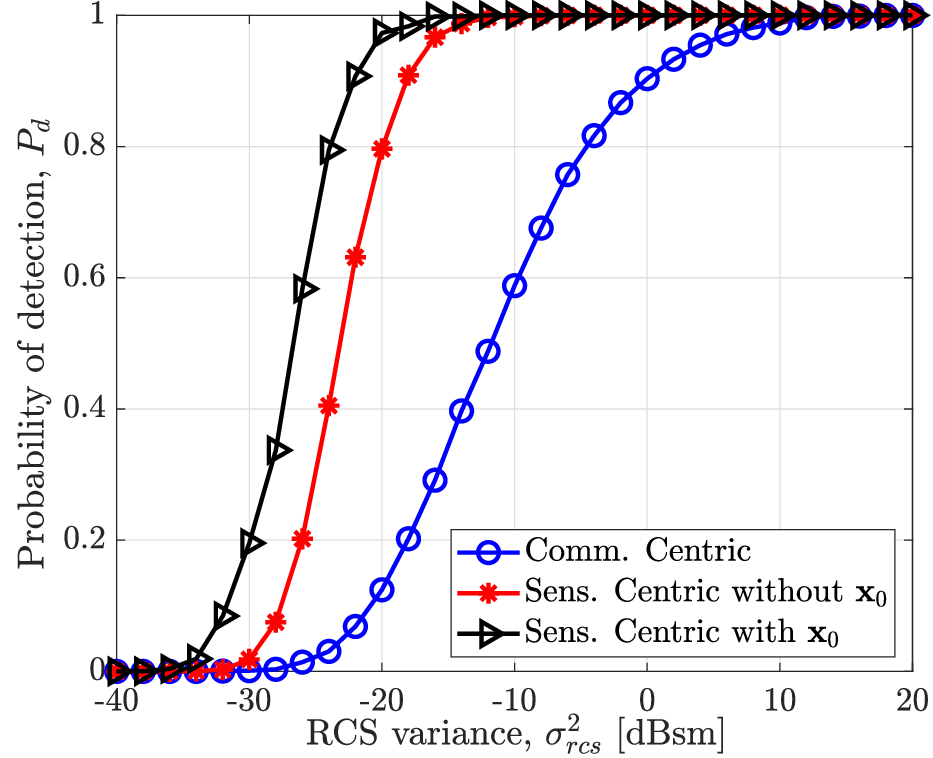}
         \caption{}
         \label{pd-vs-rcs}
     \end{subfigure}
     \hfill
     \begin{subfigure}[b]{0.329\textwidth}
         \centering
         \includegraphics[width=\textwidth]{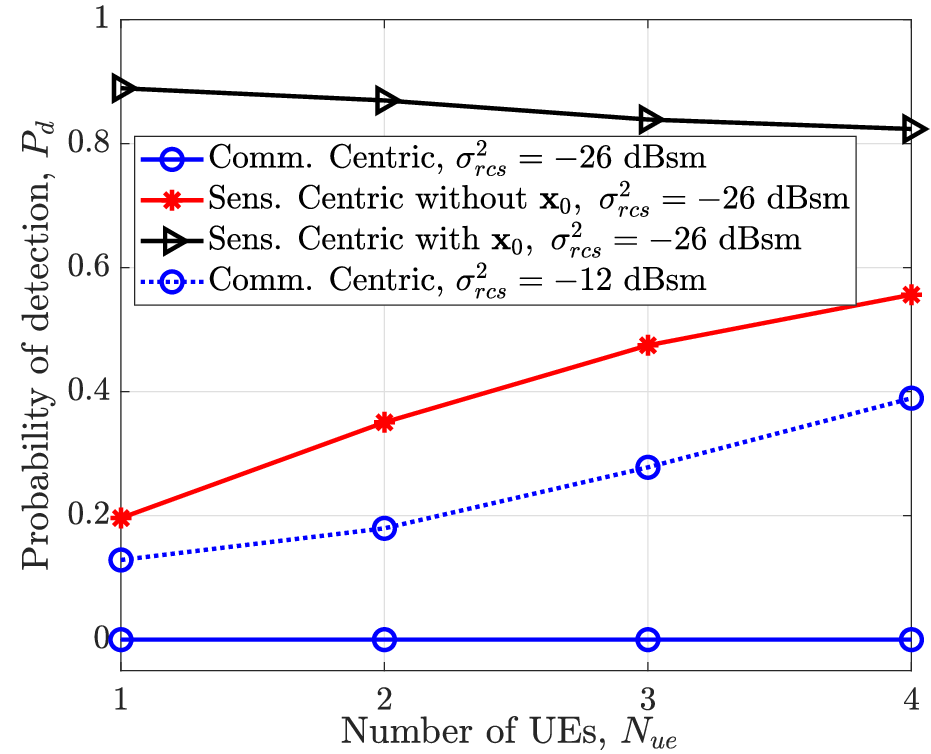}
         \caption{}
         \label{pd-vs-nue}
     \end{subfigure}
     \hfill
     \begin{subfigure}[b]{0.329\textwidth}
         \centering
         \includegraphics[width=\textwidth]{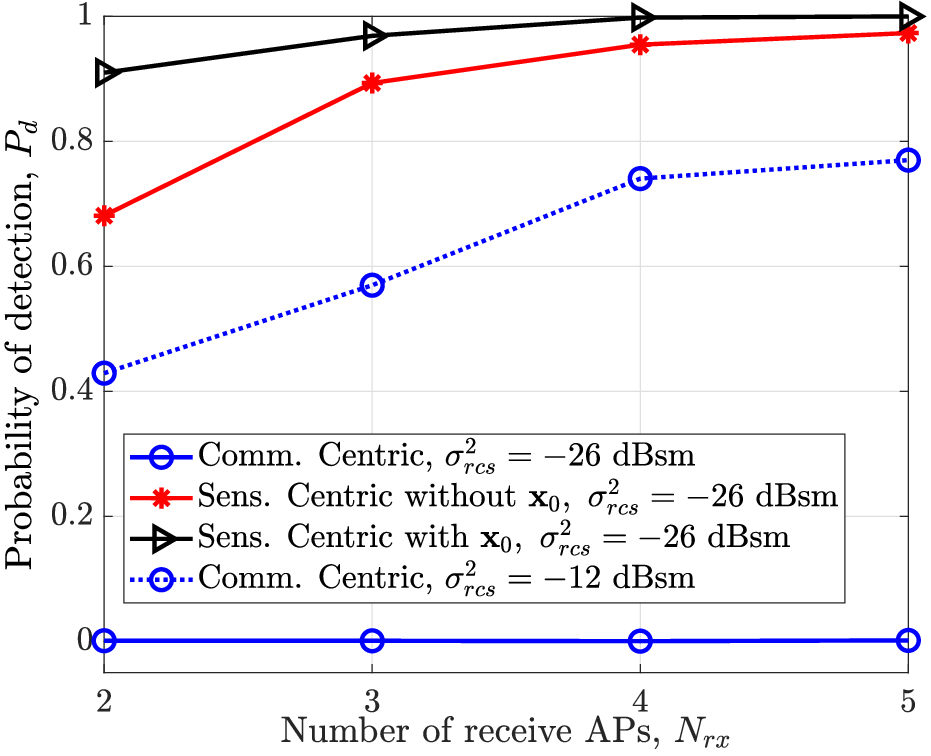}
         \caption{}
         \label{pd-vs-nrx}
     \end{subfigure}
        \caption{\small Target detection probability $\left(P_d\right)$ versus (a) variance of the target's RCS $\left(\sigma_{\rcs}^2\right)$, for $N_{rx} = 2,~ N_{ue} = 2$; (b)  Number of UEs $\left(N_{ue}\right)$, for $N_{rx} = 2$; (c)  Number of receive APs $\left(N_{rx}\right)$, for $N_{ue} = 2$. For all plots, $M = 16,~ N = 8,~ L=4,~ N_{tx} = 8, P_{fa} = 10^{-5}$.}
        \label{fig:three-graphs}
        \vspace{-0.3cm}
\end{figure*}

We compare the target detection performance of the proposed sensing-centric approach which allocate power to maximizes the sensing SNR subject to \eqref{soc-uethresh} and \eqref{soc-budget}, against a baseline communication-centric scheme that allocates power to minimize the total power consumed at the transmit APs subject to the same constraints. The latter is denoted as ``\emph{Comm. Centric}" in the legends of the simulation results. In the sensing-centric approach, we consider two variants: $(i)$ without dedicated sensing symbols $\vec{x}_0$ denoted as ``\emph{Sens. Centric without $\vec{x}_0$}", and $(ii)$ with dedicated sensing symbols $\vec{x}_0$ denoted as ``\emph{Sens. Centric with $\vec{x}_0$}". 

In Fig. \ref{pd-vs-rcs}, we plot the probability of target detection $\left(P_d\right)$ as a function of the variance of the target's RCS $\left(\sigma_{\rcs}^2\right)$. As expected, for the same communication QoS and power budget constraints, the sensing-centric schemes for ISAC-OTFS perform better than the baseline communication-centric approach. Particularly, the sensing-centric approach with dedicated sensing symbols leads to better target detection probability in comparison to the sensing-centric approach without sensing symbols $\mathbf{x}_0$. 
As can be seen from Fig. \ref{pd-vs-rcs}, the target detection probability of the communication-centric scheme is close to zero even when $\sigma_{rcs}^2 \approx -20$ dBsm, while the sensing-centric schemes with and without sensing symbols $\vec{x}_0$ lead to a probability of detection close to 1 and 0.8 respectively for the same value of $\sigma_{rcs}^2$. Consequently, for better clarity, in the remainder of the paper, we plot the target detection performances of the communication-centric scheme for $\sigma_{rcs}^2 = -26$ dBsm and $\sigma_{rcs}^2 = -12$ dBsm. However, we plot the performance of the sensing-centric schemes only for $\sigma_{rcs}^2 = -26$ dBsm since for $\sigma_{rcs}^2 = -12$ dBsm they result in $P_d = 1$ as shown in Fig. \ref{pd-vs-rcs}. 

In Fig. \ref{pd-vs-nue}, we show the impact of the number of UEs on the probability of target detection. As the number of UEs increases, $P_d$ improves when only communication symbols are used for target detection, except in the baseline communication-centric case with $\sigma_{\textsf{rcs}}^2 = -26$ dBsm because it has zero detection probability as shown in Fig. \ref{pd-vs-rcs}. This is because as the number of UEs increases, the chances of more UEs being close to the target increases. Then, the objective in \eqref{opti} allocates more power to those UEs so that the reflections off the target are stronger, thereby leading to better detection probability. When dedicated sensing symbols are used in addition to the communication symbols, the probability of detection gradually goes down with an increase in the number of UEs. This is because the power allocated to the sensing symbols goes down to maintain the communication QoS constraints in \eqref{soc-uethresh}. 

Fig. \ref{pd-vs-nrx} shows that the target detection probability $P_d$ increases with the number of receive APs $N_{rx}$, excluding the baseline communication-centric case with $\sigma_{\textsf{rcs}}^2 = -26$ dBsm because it has zero detection probability as shown in Fig. \ref{pd-vs-rcs}. The number of reflections captured from the target increases with $N_{rx}$, which leads to a better detection with an increase in $N_{rx}$. 

\section{Conclusion}
We studied the target detection performance of a cell-free MIMO ISAC-OTFS system by optimizing power allocation to maximize sensing performance while ensuring minimum required communication performance and power budget. We illustrated that the probability of target detection improves as the number of receive APs, the number of UEs, and target RCS increases. We also showed that for a given number of UEs, transmitting dedicated sensing symbols with communication signals improves the target detection performance of the cell-free MIMO ISAC-OTFS system.

\bibliographystyle{IEEEtran}
\bibliography{IEEEabrv, ISAC_OTFS_ref}

\end{document}